\documentclass[]{aastex}
\usepackage{spr-astr-addons}
\usepackage{url}
\usepackage{graphicx}
\usepackage{epstopdf}
\usepackage[fleqn]{amsmath}
\RequirePackage{color}

\begin{document}

\title{Shaping the solar wind temperature anisotropy by the interplay of electron and proton 
	instabilities}

\slugcomment{Not to appear in Nonlearned J., 45.}
\shorttitle{Electromagnetic ion cyclotron instability}
\shortauthors{Shaaban et al.}

\author{S.~M.~Shaaban\altaffilmark{1,2}} \and \author{M.~Lazar \altaffilmark{1,3}} \and \author{S.~Poedts \altaffilmark{1}} \and \author{A.~Elhanbaly\altaffilmark{2}}

\altaffiltext{1}{Centre for Mathematical Plasma Astrophysics, Celestijnenlaan 200B, B-3001 Leuven, Belgium.} \email{shaaban.mohammed@student.kuleuven.be}
\altaffiltext{2}{Theoretical Physics Research Group, Physics Department, Faculty of Science, Mansoura University, 35516 Mansoura, Egypt.}
\altaffiltext{3}{Institut f\"ur Theoretische Physik, Lehrstuhl IV: Weltraum- und
                Astrophysik, Ruhr-Universit\"at Bochum, D-44780 Bochum, Germany.}

\begin{abstract}
A variety of nonthermal characteristics like kinetic, e.g., temperature, anisotropies and suprathermal populations (enhancing the high energy tails of the velocity distributions) are revealed by the in-situ observations in the solar wind indicating quasistationary states of plasma particles out of thermal equilibrium. Large deviations from isotropy generate kinetic instabilities and growing fluctuating fields which should be more efficient than collisions in limiting the anisotropy (below the instability threshold) and explain the anisotropy limits reported by the observations. The present paper aims to decode the principal instabilities driven by the temperature anisotropy of electrons and protons in the solar wind, and contrast the instability thresholds with the bounds observed at 1~AU for the temperature anisotropy. The instabilities are characterized using linear kinetic theory to identify the appropriate (fastest) instability in the relaxation of temperature anisotropies $A_{e,p} = T_{e,p,\perp}/ T_{e,p,\parallel} \ne 1$. The analysis focuses on the electromagnetic instabilities driven by the anisotropic protons ($A_p \lessgtr 1$) and invokes for the first time a dynamical model to capture the interplay with the anisotropic electrons by correlating the effects of these two species of plasma particles, dominant in the solar wind.
\end{abstract}


\section{Introduction}

Due to a continuous presence of observational missions in space, the solar wind is currently 
exploited as a natural laboratory for studying the plasma mechanisms and effects including 
kinetic instabilities driven locally by the temperature anisotropy of plasma particles. The 
existence and viability of these mechanisms of instability are confirmed by a recent combined 
analysis of the plasma particle distributions and the enhanced wave fluctuations observed in the 
solar wind, see \cite{Gary2016}. Instead, the back effects of the growing fluctuating fields 
scattering the plasma particles and limiting their anisotropy remain controversial, especially 
the role played in the relaxation process by the cyclotron electromagnetic fluctuations usually 
dominating the direction parallel to the stationary magnetic field \citep{Kasper2003, 
Hellinger2006, Bale2009}. 

Driven by an excess of perpendicular temperature $T_{\perp} > T_{\parallel}$ cyclotron modes grow faster
than other instabilities like mirror instability which may develop in the same conditions. 
Resonant interactions, i.e., cyclotron resonance with plasma particles, are therefore expected to 
be effective in the relaxation of temperature anisotropy in this case. 
The interest to explain these effects and their consequences in the solar wind has been 
boosted after the simulations suggested that the mirror instability is not an 
effective pitch angle scatterer of protons \citep{Mckean1992,Mckean1994,Gary1993a}
and the cyclotron anisotropy instability may be assumed the primary mechanism to 
constrain the proton anisotropy \citep{Gary1994Correlation}. Further confirmation 
of this constraint can also be obtained in a straightforward way, namely, by fitting 
the instability thresholds predicted by the linear theory with the anisotropy 
limits reported by the observations (see explanatory arguments and references 
in \cite{Gary1994Limited}). However, the instability thresholds derived from 
simplified models assuming plasma particles bi-Maxwellian distributed and 
minimizing the effects of electrons considering them isotropic, do not 
provide a good agreement with the observations. Thus, thresholds of the 
electromagnetic ion (proton) cyclotron (EMIC) instability simply do not 
align to the limits of the proton temperature anisotropy in the solar wind, 
but are markedly lower than these limits, which instead appear to be better 
described by the thresholds of the mirror (aperiodic) modes instability 
\citep{Hellinger2006}. In the opposite case, an excess of parallel temperature 
$T_{\perp} < T_{\parallel}$ may ignite two branches of firehose instability, 
one destabilizing the electromagnetic cyclotron modes propagating mainly
in the parallel direction but with an opposite polarization, i.e., 
right-handed (RH) if driven by protons and left-handed (LH) if driven by 
electrons, and the other one destabilizing highly oblique and aperiodic modes.
Again, when the firehose thresholds are derived with simplified models the 
temperature anisotropy in the solar wind is better constrained by the aperiodic 
instability \citep{Hellinger2006}, which, however, cannot undergo cyclotron 
resonant interactions with plasma particles. In both these two cases, the anisotropy
thresholds derived for the instabilities of cyclotron modes lie below the 
anisotropy limits reported by the observations in the solar wind, seeming 
that the instability thresholds may be overestimated by using simplified models 
for the velocity distributions of plasma particles and neglecting their 
interplay.
Two distinct classes of mechanisms may be at work in the solar wind generating temperature anisotropies of 
plasma particles. These are either the large scale mechanisms like adiabatic expansion (leading to 
$A = T_{\perp} / T_{\parallel} < 1$) and magnetic compression (leading to $A > 1$), or the small scale
heating and acceleration of plasma particles by their resonant interactions with electromagnetic fields fluctuations. 
Large scale mechanisms act in the same manner on both species, electrons and protons (subscripts $e$ and 
$p$, respectively), expecting to provide a direct correlation between their anisotropies measured in the solar 
wind, i.e., both species with $A_{e,p} > 1$, or both with $A_{e,p} < 1$. 
Binary collisions are not efficient enough to reduce the anisotropy and affect this correlation
of the electron and proton anisotropies, but the small scale mechanisms like the wave-particle interactions, 
usually conditioned by the presence of different wave fluctuations may accelerate plasma particles preferentially,
e.g., in direction perpendicular to the magnetic field by the cyclotron resonance (leading to $A > 1$), 
or in parallel direction by the Landau (transit-time) damping (leading to $A < 1$). Moreover, the electrons 
mainly resonate with the high-frequency waves while the protons react to the low-frequency modes. An 
anti-correlation between the electron and proton anisotropies, i.e.,  $A_{e} \gtrless 1$ and  $A_{p} 
\lessgtr 1$, can therefore result from local mechanisms involving either microinstabilities or 
damping of small scale fluctuations.

A quantitative analysis with systematic evidences and estimations of these correlations between 
the electron and proton anisotropies in space plasmas is not reported yet, at least to our knowledge,  
but some qualitative elements can however be extracted. 
Thus, an implication of the large scale mechanisms in generating temperature anisotropy of plasma particles 
seems to be confirmed by the observations, which show a radial evolution of the temperature anisotropy from an exclusive 
$A = T_{\perp} / T_{\parallel} > 1$ at low heliocentric distances $ \sim 0.3$~AU \citep{Matteini2007}, 
where the interplanetary magnetic field is more intense, to a dominant $A<1$ after the expansion at large radial distances $\sim 1$~AU
\citep{Kasper2003,Stverak2008}. On the other hand, the anisotropy-driven instabilities
enhance the small-scale fluctuations, which may play two distinct roles, either to maintain the 
anisotropy correlation of the electrons and protons, e.g., by cyclotron electromagnetic instabilities,
or to transfer the free energy between plasma species, e.g., by firehose instability, and eventually induce an 
anti-correlation of their anisotropies. Indeed, radial profile of the temperature anisotropy departs from CGL predictions \citep{Matteini2007} suggesting perpendicular heating by wave turbulence, while a a parallel cooling may likely be related to microinstabilities connected with the structure of the proton velocity distribution function \citep{Hellinger2013}.

The present paper presents an advanced description of these instabilities on the basis of a refined and,
tentatively, more realistic model, which takes into account the effects of suprathermal electrons, 
ubiquitous in the solar wind, as well as different couplings between the electron and proton anisotropies. 
The dispersion formalism is provided in Sec.~2 on the basis of a Vlasov kinetic 
approach for a plasma of electrons and protons described by (bi-)Kappa distribution functions. 
Kappa power-laws are generalized models empirically introduced to describe with accuracy 
the velocity distributions measured in space plasmas \citep{Vasyliunas1968, Maksimovic1997, 
Christon1989}. Standard (bi-)Maxwellian models can reproduce only the low-energy core of the 
measured distributions, while (bi-)Kappa can also incorporate the high-energy tails of the 
distributions, which are markedly enhanced by the suprathermal populations. The unstable 
solutions are discussed in Sec.~3 for different situations mainly conditioned by the 
interplay of anisotropic electrons and protons. We focus on the electromagnetic instabilities
driven by the anisotropic protons and invoke for the first time a dynamical model 
to include the effects of anisotropic electrons. In Sec.~4 we contrast the instability 
thresholds with the observations, and perform a comparative analysis with the previous 
results from simplified approaches. Conclusions are presented in Sec.~5.

\section{Dispersion--stability formalism: transverse modes}

For a collisionless and homogeneous electron--proton plasma, the electromagnetic modes in a 
direction parallel to the stationary magnetic field (${\bf k} \parallel {\bf B}$) decouple 
from the electrostatic oscillations, and their instabilities may display maximum 
growth rates, e.g., cyclotron instabilities \citep{Kennel1966}. Provided by a linear 
Vlasov-Maxwell dispersion formalism \citep{Krall1973}, the dispersion relations for these electromagnetic modes read
\begin{equation}
     \begin{aligned}
1+\sum_{\alpha=e,p}\frac{\omega _{p,\alpha}^{2}}{\omega ^{2}}\left[\frac{\omega}{k\; u_{\alpha,\parallel}} Z_{\alpha,\eta}\left( \xi _{\alpha,\eta}^{\pm }\right)+ \left( A_{\alpha}-1\right)\right.\\
\left.\times\left\{ 1+\xi _{\alpha,\eta}^{\pm} \; Z_{\alpha,\eta}\left(\xi _{\alpha,\eta}^{\pm }\right) \right\} \right]=\frac{c^{2}k^{2}}{\omega ^{2}},
\label{e1}
      \end{aligned}
\end{equation}
where $\omega$ is the wave-frequency, $k$ \ is the wave-number, $c$ is the speed of light, 
$\omega_{p,\alpha}^{2}=~4\pi n_{\alpha}e^{2}/m_{\alpha }$ are the plasma
frequencies for protons (subscript $\alpha=p$) and electrons (subscript $\alpha=e$),  
$A_{\alpha}=T_{\alpha, \perp}/T_{\alpha, \parallel}$ are the temperature anisotropies, 
$\pm $ denote the circular polarizations, right-handed (RH) and left-handed (LH), respectively. 
$Z_{\alpha,\eta}\left( \xi _{\alpha,\eta}^{\pm }\right)$ may denote either the plasma dispersion 
function for (bi)-Maxwellian (subscript~$\eta=M$) distributed plasmas \citep{Fried1961}, or the 
modified dispersion function for Kappa (subscript $\eta=~\kappa$) distributed plasmas as derived in 
\cite{Lazar2008}, and $u_{\alpha,\parallel}$ are the corresponding thermal thermal velocities 
\citep{Lazar2015Destabilizing}. See appendix A for the explicit definitions of these quantities.

Since the presence of anisotropic ($A_e \ne 1$) electrons can change the dispersive properties 
of the electromagnetic modes, including those destabilized by the proton anisotropy $A_p \ne 1$ 
\citep{Kennel1968,Lazar2011,Michno2014,Shaaban2015,Shaaban2016Supra}, here we propose a dynamical 
model for the interplay of the electrons and protons by correlating their temperature anisotropies 
$A_{e,p}\neq1$, as well as their plasma parallel betas $\beta_{e,p,\parallel}$. Previous studies 
carried out by \cite{Kennel1968,Lazar2011,Michno2014,Shaaban2015,Shaaban2016Supra} have assumed
constant values of the electron anisotropy $A_e$ and plasma beta $\beta_{e,\parallel}$, independent
of proton properties. On the other hand, our present analysis includes the effects of the suprathermal 
populations of electrons, which are ubiquitous in the solar wind.

In order to proceed and make the analysis more transparent, we rewrite the linear dispersion 
relation (\ref{e1}) in terms of normalized quantities
\begin{equation}
     \begin{aligned}
\tilde{k}^{2}=A_{p}-1+\frac{ A_{p}\left(\tilde{\omega}\pm 1\right)\mp 1}{\tilde{k}\sqrt{\beta _{p,\parallel}}}Z_{p, \eta}\left( \frac{\tilde{\omega}\pm1}{\tilde{k}\sqrt{\beta_{p,\parallel}}}\right)\\
+\mu \left({A_e}-1\right)+\mu \frac{{A_e}\left( \tilde{\omega}\mp \mu\right)\pm \mu}{\tilde{k}\sqrt{\mu \beta_{e,\parallel}}}Z_{e,\eta}\left( \frac{\tilde{\omega}\mp \mu }{\tilde{k}\sqrt{\mu \beta _{e,\parallel}}}\right)
    \label{e2}   
     \end{aligned}     
\end{equation}
where $\tilde{\omega}=\omega /\Omega _{p}$, $\tilde{k}=~kc/\omega_{p,p}$, $\mu =~m_{p}/m_{e}$ is the
proton/electron mass ratio, $\beta_{\alpha, \parallel} = 8 \pi n_{e} k_{B} T_{\alpha,\parallel} / B^{2}$ are 
the parallel plasma betas for protons (subscript $\alpha=p$) or electrons (subscript $\alpha=e$).
\textbf{Now we assume that the electron and proton temperature anisotropies are correlated}
\begin{align}
A_e=~A_p^{\delta}
\end{align}
\textbf{with a correlation index $\delta$, eventually indicated by the observations.}
The  correlation index may be either positive $\delta >~0$ when both species have the same type 
of anisotropy with respect to the stationary magnetic field ${\bf B}$, i.e., if $A_p \gtrless 1$ then $A_e \gtrless 1$, 
or negative $\delta < 0$, when electrons and protons have opposite anisotropies, i.e., if $A_p\gtrless1$ then $A_e\lessgtr1$. 
In the following calculations our reference is the classical case with isotropic electrons, i.e., $A_e=1$, that here is
obtained for a correlation index $\delta =0.0$.

\section{Unstable solutions: interplay of protons and electrons}

In this section we examine the electromagnetic instabilities driven by proton anisotropies 
$A_p\neq1$ under the influence of anisotropic electrons by means of the correlation-index 
$\delta$. For the electrons the velocity distributions measured in space plasmas
may be considerably  enhanced by the suprathermal populations and we therefore consider them 
(bi-)Kappa distributed. The proton data invoked in our analysis is measured by SWE/WIND 
\citep{Ogilvie1995} with velocities corresponding to a kinetic energy in the range of 
$150$~eV to $8$~keV excluding suprathermal populations. 
In general, the presence of suprathermal protons is indeed less significant and we can assume 
the protons to be more thermalized and bi-Maxwellian distributed. 

We divide our analysis into two distinct classes of instabilities according to the proton anisotropy. First 
we consider the instabilities developed by an excess of parallel temperature, i.e., $A_p > 1$.  
Among these, the firehose instability is the fastest growing mode and is RH circularly polarized 
when is propagating in direction parallel to the magnetic field. In the opposite situation,
when protons exhibit an excess of perpendicular temperature, i.e., $A_p < 1$, the fastest developing 
instability is that of the electromagnetic ion cyclotron (EMIC) instability propagating 
in direction parallel to the magnetic field and with a LH circular polarization.

\subsection*{\textbf{3.1 Solar wind protons with $T_{p,\parallel}>T_{p,\perp}$}}

 \begin{figure}[t]
   \centering
   \includegraphics[width=7.5cm,trim={ 0  0.cm 0cm 0.0cm},clip]{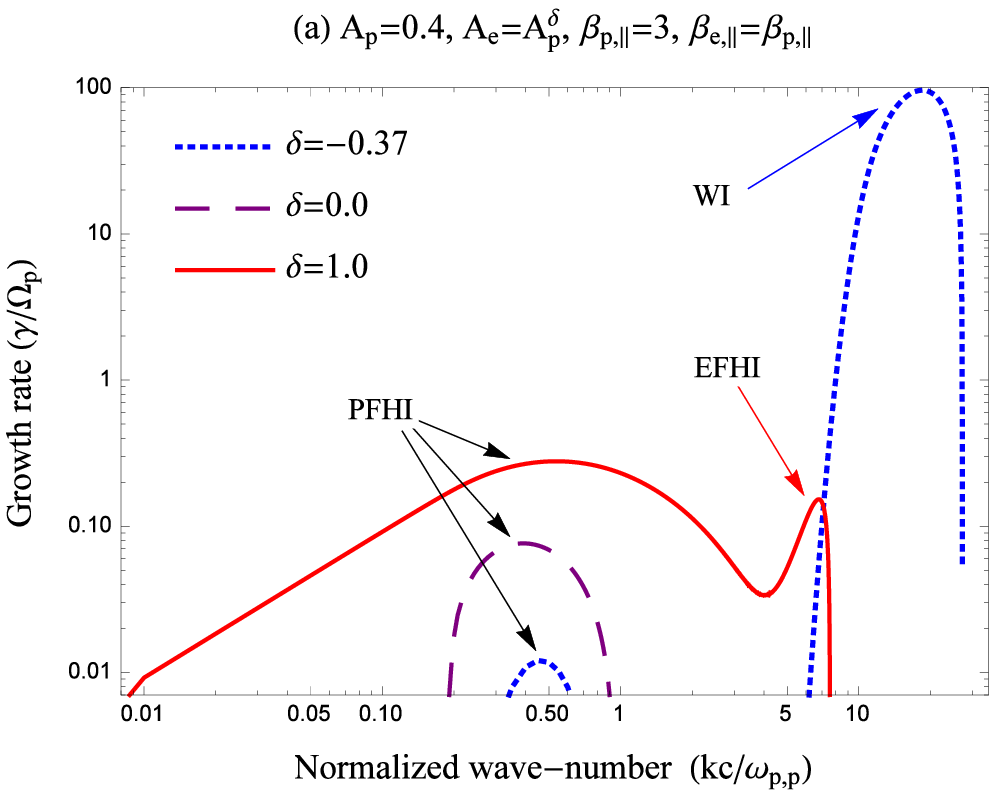}
      \includegraphics[width=7.5cm,trim={0 0 0.1cm 0.4cm},clip]{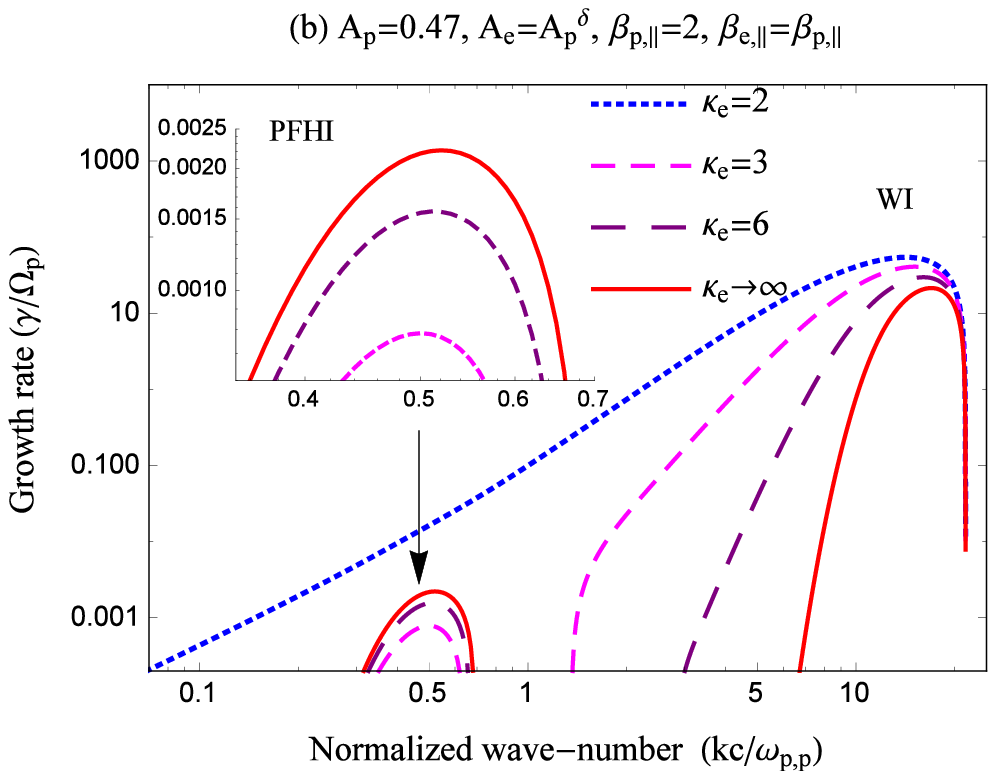}
      \caption{Effects of $\delta-$index $=-0.37,0.0,1.0$ (top), and the $\kappa-$~index=$2,3,6,\infty$ 
			with $\delta=-0.3$ (bottom) on the growth rates of PFHI instability. The plasma parameters are explicitly given in each panel.}
         \label{f1}
   \end{figure}
 \begin{figure}[t]
   \centering
      \includegraphics[width=7.6cm]{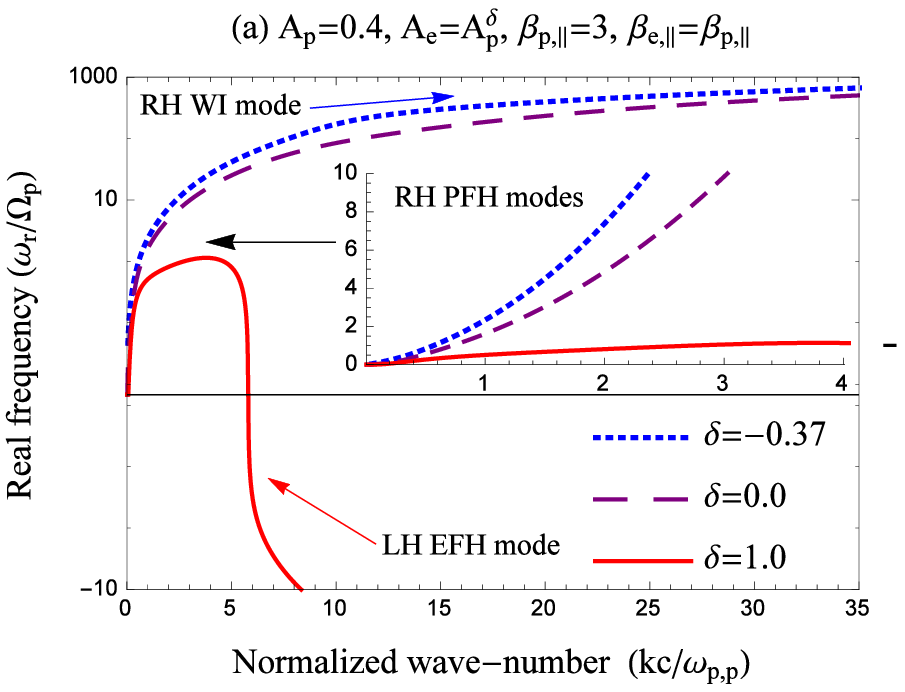}
      \includegraphics[width=7.4cm]{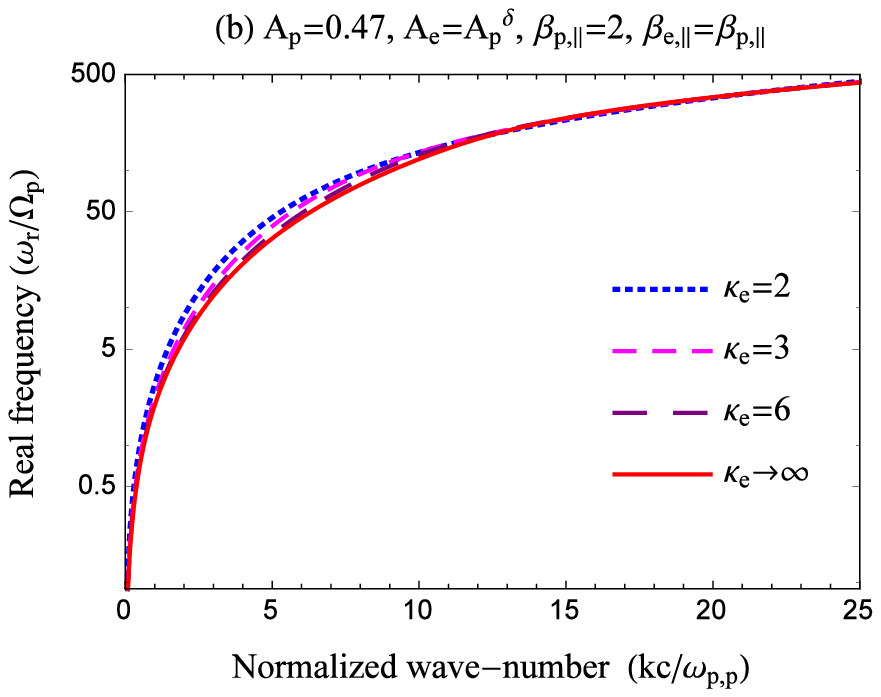}
      \caption{Wave-frequencies of the unstable modes in Fig. 1. The plasma parameters are 
			mentioned in each panel.}
         \label{f1r}
   \end{figure}
The effects of the new correlation-index $\delta$ on the growth rates of the PFHI 
are displayed in Figure~\ref{f1}~(a). The unstable solutions are computed, for $A_p=~0.4$, 
$\beta_{e,\parallel}=~ \beta_{p,\parallel}=3$, and for different values of $\delta=
1.0$, $0.0$, $-0.37$ (implying different electron anisotropies, respectively, $A_e=0.4, 
1.0, 1.4$). The growth rates display two distinct peaks when the electrons are anisotropic, 
i.e., $\delta=$~1.0, -0.37. The first peak at low wavenumbers corresponds to the PFHI, 
while the second peak may represent either the electron FHI (EFHI), see the solid (red) line, 
when the correlation-index $\delta=1$, or the whistler instability (WI), see the dotted (blue) line, 
when the correlation-index $\delta=-0.37$. Driven by an electron anisotropy $A_e<1$ the EFHI 
propagating parallel to the magnetic field is LH circularly polarized and has a frequency in 
the range of $\omega_p<\omega_r\ll\vert\Omega_e\vert$, while the WI is a RH circularly polarized 
mode destabilized by the anisotropic electrons with $A_e>1$ and with the frequency $\omega_p<
\omega_r<\vert\Omega_e\vert$. Results in Figure~\ref{f1}~(a) show that the growth rates of the PFHI are inhibited by anisotropic electrons described by a negative $\delta-$index (implying $A_e>1$), while the same growth rates are stimulated by a positive $\delta-$index (implying $A_e<1$). Physically,
under the inhibiting effect of a negative $\delta <0$ on the PFHI, the instability needs higher 
values of the anisotropy $A_p$ or higher values of the parallel plasma beta $\beta_{p,\parallel}$ 
to achieve the same growth rate. We should therefore expect that the PFHI thresholds will 
move towards higher values of plasma beta $\beta_{p,\parallel}$, and shape better 
the anisotropy limits observed in the solar wind (see Sec.~4). For the opposite case, when 
$\delta>0$, the instability is stimulated and threshold conditions should diminish towards 
lower values of plasma beta $\beta_{p,\parallel}$. 

Based on these premises, the effect of the suprathermal electrons is shown in Figure \ref{f1}~(b) 
only for a negative correlation-index $\delta=-0.3$. The growth rates are computed 
for $A_p=0.47$, $\beta_{e,\parallel}=\beta_{p,\parallel} = 2$, and for different values of the 
power-index $\kappa_e=2, 3, 6, \infty$. The PFHI growth rates are suppressed by increasing the 
suprathermal population, i.e., lowering $\kappa$, see the subplot (zoomed) in Figure \ref{f1}~(b), 
while the WI growth rates are enhanced. In other words, the inhibiting effect of the anisotropic
electrons described by a negative $\delta <0$ on the PFHI is enhanced by the suprathermal electrons.
Moreover, this instability may be completely suppressed with decreasing the power-index, e.g.,  for 
$\kappa_e=2$, only the WI develops (blue dotted line) at electron scales. Note that 
the growth rate of the WI obtained for $A_p=0.47$ cannot ditinguish from that obtained with 
isotropic protons $A_p=1$ (not shown here). 
These new results apparently contradict those obtained by \cite{Lazar2011} in a 
study of the PFHI cumulatively driven by the anisotropic protons $A_p<1$ and electrons $A_e>1$ 
(see Figure 5 in \cite{Lazar2011}). In their study the WI growth rates were found to decrease with 
increasing the electron suprathermal population as a result of a different Kappa approach with 
a $\kappa-$independent temperature that recently was proven inappropriate for such an 
analysis \citep{Lazar2015Destabilizing}. More results on the kinetic instabilities using 
approaches with a $\kappa-$dependent temperature, as used in the present paper, 
can be found in \cite{Leubner2000, Leubner2001, Lazar2015Destabilizing}, and \cite{Shaaban2016Supra}.

\textbf{For a complete picture, Figure \ref{f1r} displays the real frequencies corresponding 
to the unstable solutions in Figures \ref{f1}. When the anisotropies of protons and electrons are correlated
by a positive $\delta=1$, i.e., $A_e < 1$, the wave-frequencies in panel~(a) confirm a conversion of
the RH-polarized PFH modes to the LH-polarized EFH modes (solid red line) by changing the sign in between the 
PFHI and EFHI peaks. At these low frequencies, the LH and Rh branches are relatively close to each other 
making possible a conversion, determined in this case by the interplay of the PFH and EFH instabilities. 
Otherwise, the RH branch (which is destabilized at low frequencies by the PFHI) extends smoothly (monotonically
increasing dispersion) to electron scales (dashed and dotted lines), where the anisotropic electrons with 
an anti-correlated anisotropy $A_e > 1$, as given by a negative $\delta=-0.37$, may drive the instability of
whistler modes (WI with dotted line).
In panel~(b) we show that wave-frequencies corresponding to the growth rates in Figures \ref{f1}~(b) are not 
markedly influenced by the presence of suprathermal electrons.}

\subsection*{\textbf{3.2 Solar wind protons with $T_{p,\perp}>T_{p,\parallel}$}}

\begin{figure}[t]
   \centering
   \includegraphics[width=7.5cm]{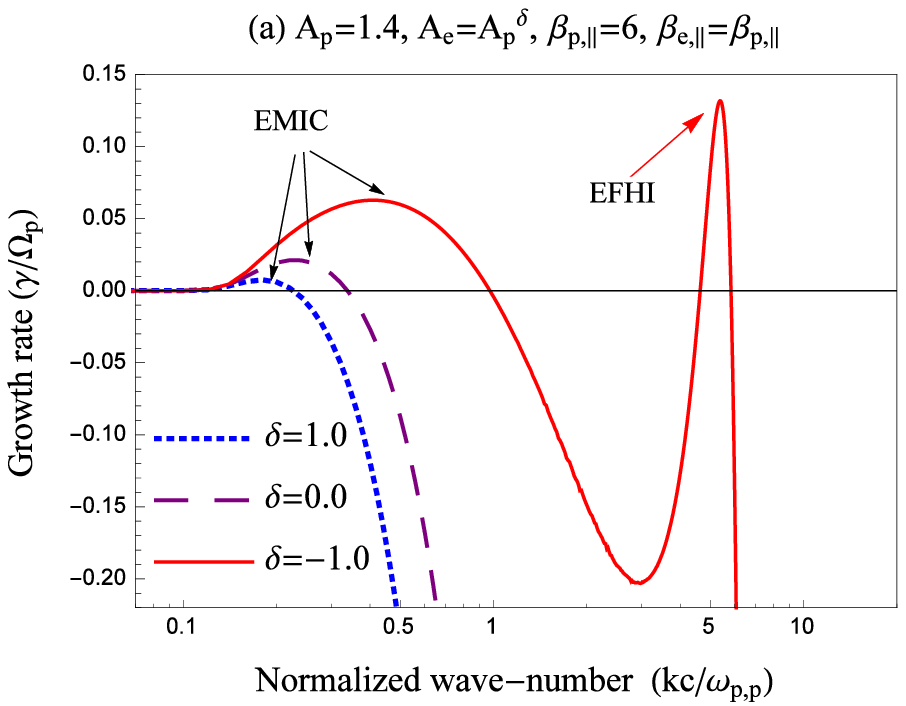}
      \includegraphics[width=7.5cm]{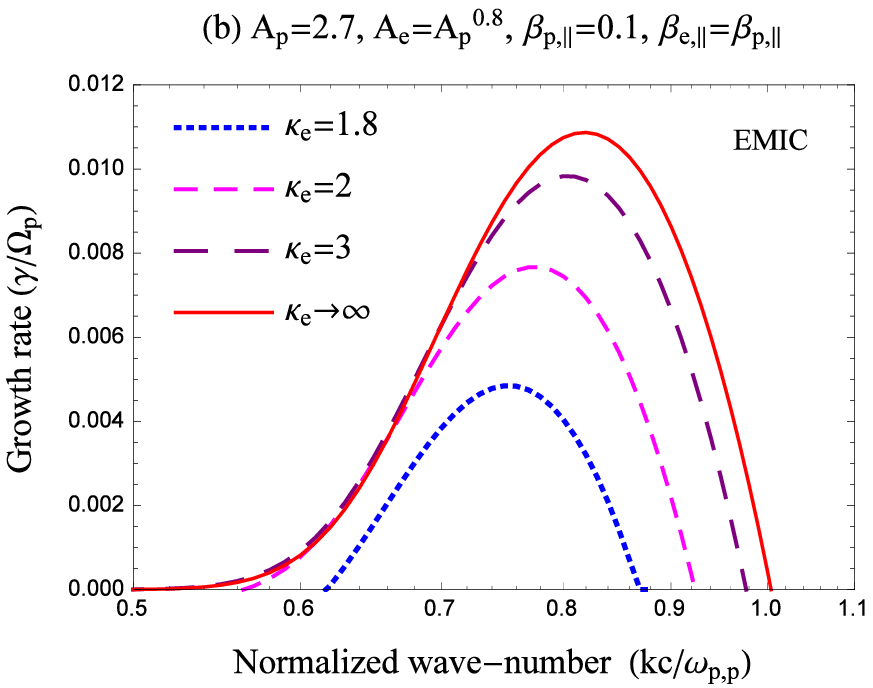}
      \caption{Effect of $\delta=1,0.0,-1$ (top) and the $\kappa$-index=$1.8,2,3,\infty$ with 
			$\delta=0.8$ (bottom) on the growth rates of EMIC instability. The plasma parameters are explicitly given in each panel.}
         \label{f2}
   \end{figure}

The electromagnetic ion cyclotron (EMIC) instability is triggered by the anisotropic protons 
with $A_p>1$ (i.e., $T_{p,\perp}>T_{p,\parallel}$). This instability develops first 
in direction parallel to the magnetic field, where the EMIC modes are LH circularly polarized.
In Figure \ref{f2}~(a) we display the growth rates of this instability driven by a temperature 
anisotropy $A_p=1.4$ for $\beta_{e,\parallel}=\beta_{p,\parallel} = 6$, and different values of
the $\delta=1.0,0.0,-1.0$, with mention that first value $\delta=1$ is carefully 
chosen to produce the same anisotropy of the electrons $A_e=1.4$ as in Figure \ref{f1}~(a). 
In this case the growth rates display two peaks only for a negative correlation-index, e.g., for 
$\delta=-1.0$ (implying $A_e=0.714$) the EMIC peak obtained at low wavenumbers is followed by 
the peak of the EFHI, see solid (red) line. For a positive $\delta > 0$, implying $A_e >1$, the WI 
is absent, as it belongs to another branch with opposite (RH) polarization and much higher frequency
($\omega_r \gg \Omega_p$). The EMIC instability is inhibited by increasing $\delta$ from negative 
values, implying protons and electrons with anti-correlated anisotropies ($A_p > 1$, $A_e < 1$) 
to positive values when these anisotropies are correlated ($A_{p,e} > 1$). 
Therefore, in Figure (\ref{f2}-b) we describe the influence of suprathermal electrons 
only for a direct correlation of the proton and electron anisotropies by a positive 
$\delta=0.8$. Growth-rates are plotted for a proton anisotropy $A_p=2.7$, the same 
plasma beta parameter for protons and electrons $\beta_{e,\parallel}=\beta_{p,\parallel}= 0.1$,
and different values $\kappa=1.8,2,3,\infty$. In the vicinity of the instability threshold 
level $\gamma_m/\Omega_p=10^{-2}$ the instability inhibits by increasing 
the suprathermal population of electrons.
 \begin{figure}[t]
   \centering
   \includegraphics[width=7.4cm]{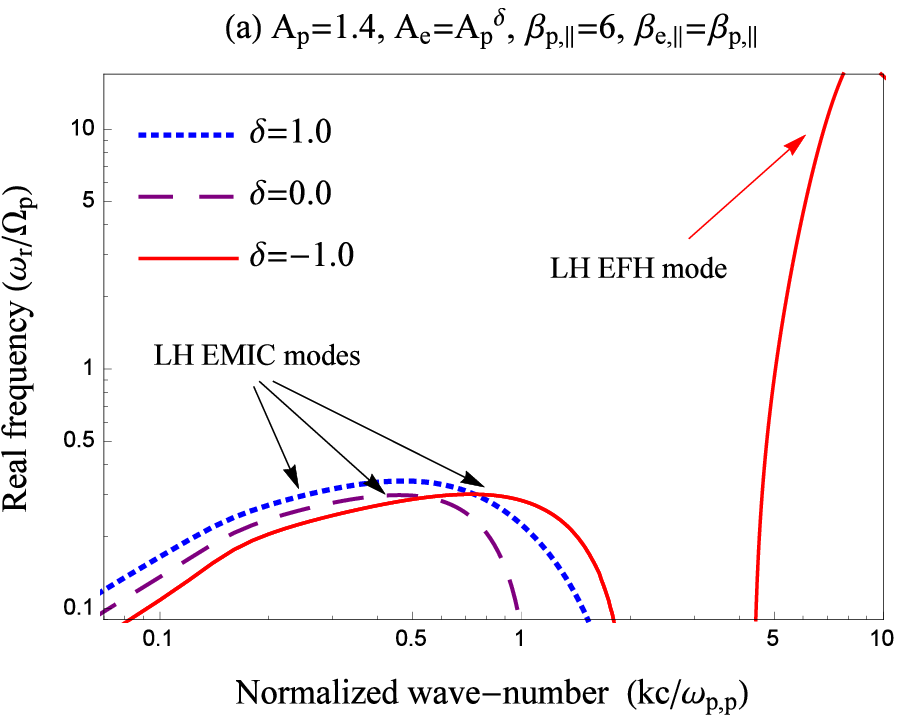}
      \includegraphics[width=7.2cm]{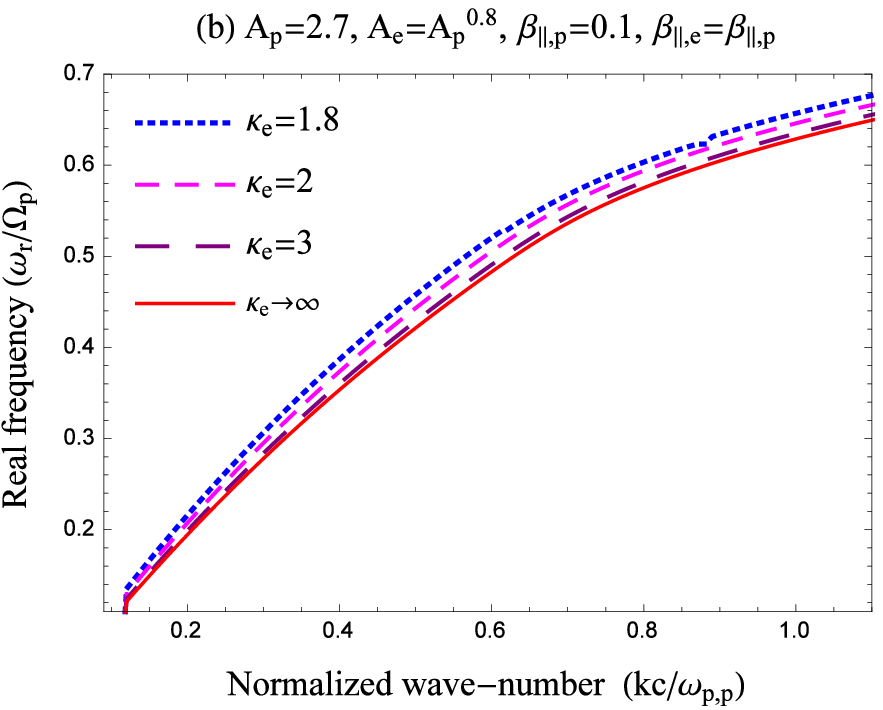}
      \caption{Wave-frequencies of the unstable modes in Fig. 3. The plasma parameters are mentioned in each panel.}
         \label{f2r}
   \end{figure}

\textbf{Figure \ref{f2r} presents the real frequencies  corresponding to the unstable 
solutions in Figure \ref{f2}. Unlike the growth rates, the wave-frequencies of the 
EMIC modes are enhanced by increasing the correlation index $\delta$, see panels~(a). 
However, the anisotropy of protons is modest (not very large) and after the EMIC 
saturation the wave-frequency changes the sign, converting to the RH branch under the 
influence of electrons, which are expected to manifest important kinetic effects in 
this case, due to their high $\beta_{e,\parallel} = 6$. The same high value of $\beta_{e,\parallel}$ 
may stimulate a LH EFHI to develop (at larger wave-numbers) when the electron anisotropy 
is anti-correlated, i.e., $\delta= -1$ (red solid line), and this is confirmed by the 
wave-frequency which becomes again positive, i.e., LH-polarized. 
In panel~(b) we show that wave-frequency of the EMIC modes are slightly enhanced by the 
presence of suprathermal electrons, but in this case the effects of electrons are minimal  
due to their small $\beta_{e,\parallel} = 0.1$.}

The results in sections~3.1 and 3.2 already suggest that instabilities thresholds can 
be enhanced, namely, by a negative correlation-index $\delta<~0$ for the PFHI and a positive 
$\delta>0$ for the EMIC instability. An analysis of these thresholds is presented 
in section~4.

\subsection*{\textbf{3.3. Insights from resonance conditions}}

\begin{figure}[t]
   \centering
 \includegraphics[width=7.2cm]{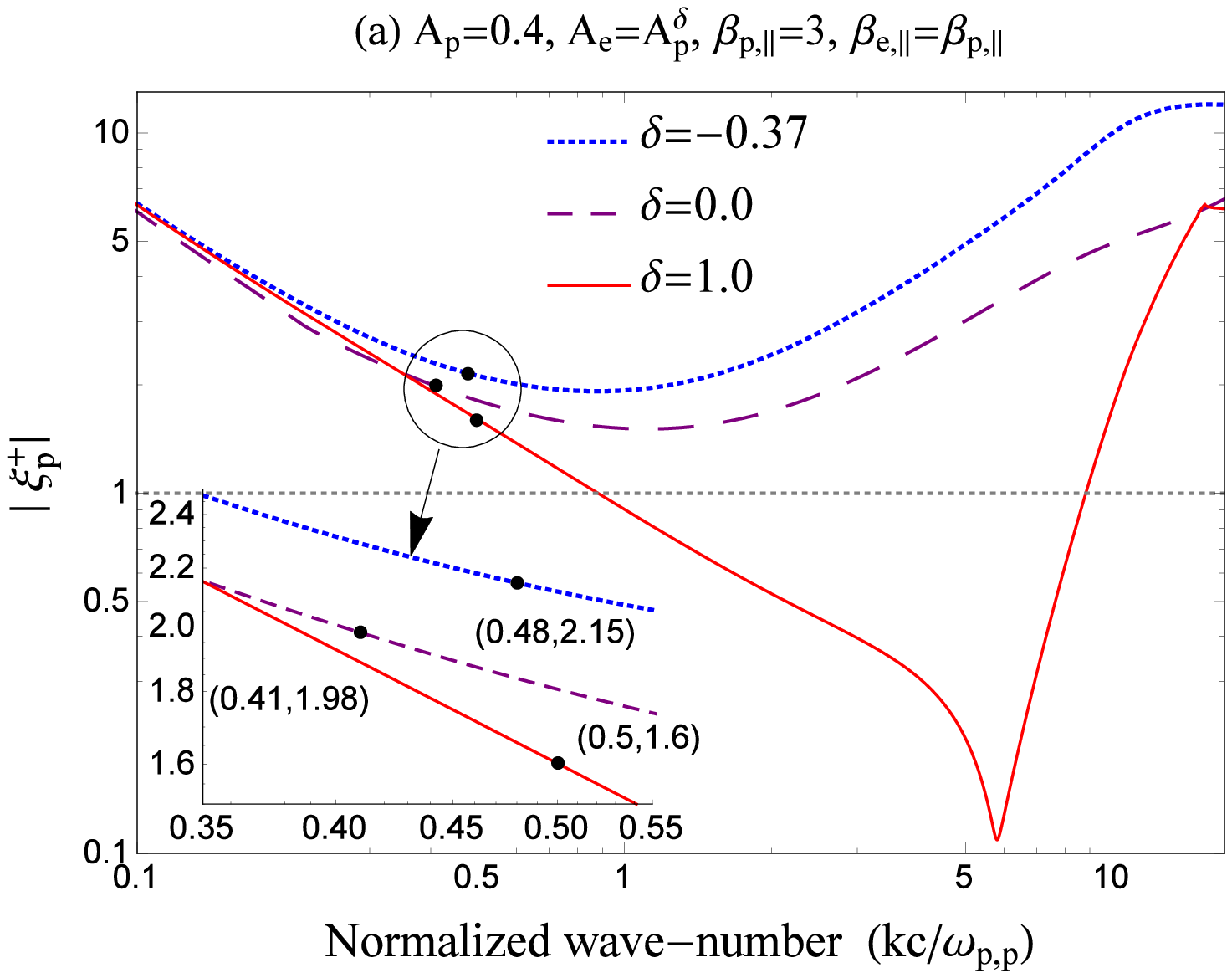} \\
  
   \includegraphics[width=7.2cm]{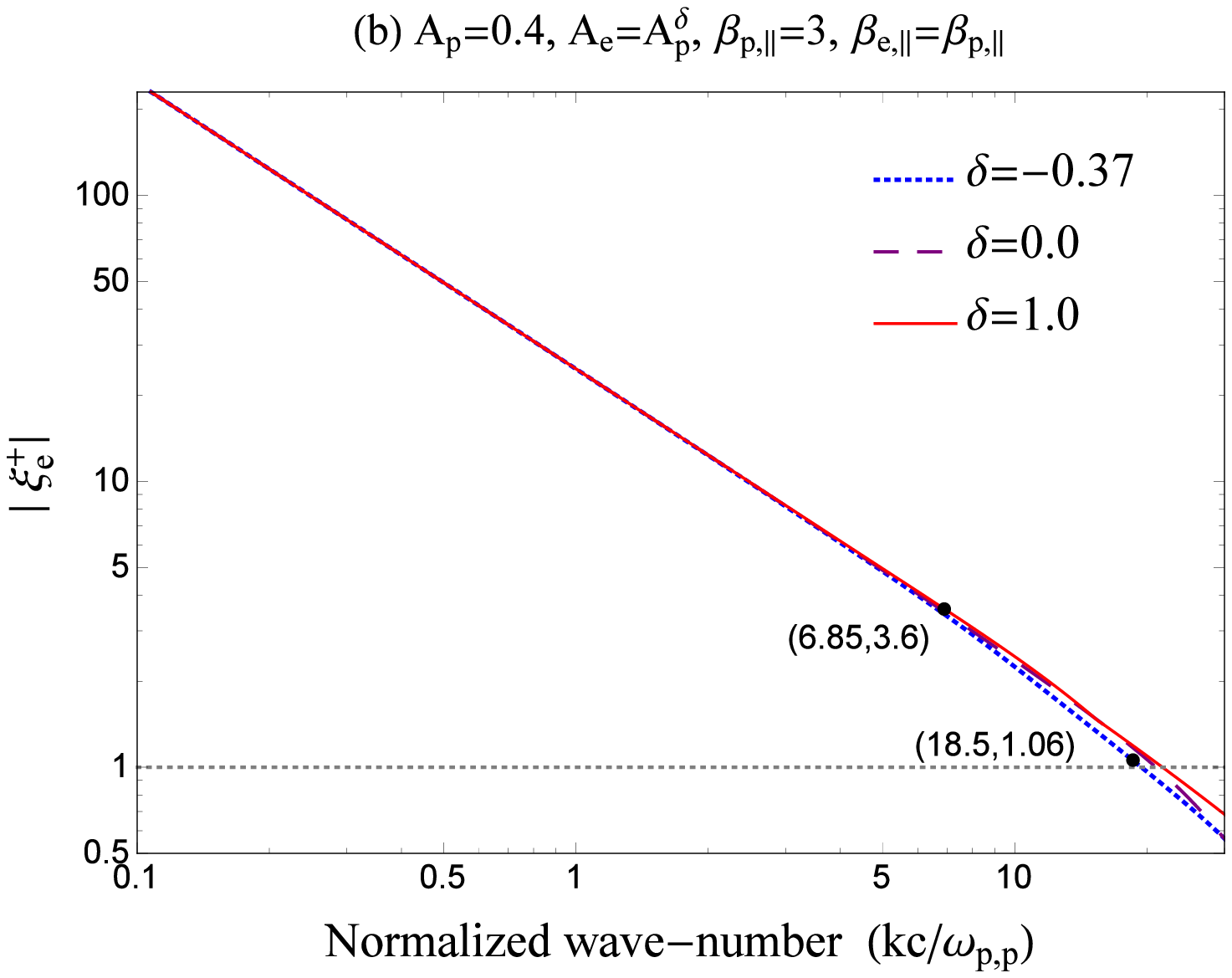}
  
      \caption{The resonant conditions for protons $\vert\xi_p^+\vert$ (top) and electrons $\vert\xi_e^+\vert$ (bottom) for the same plasma parameters in Figure (\ref{f1}-a).
              }
         \label{f3}
   \end{figure}

In these section we will try to identify the physical mechanisms behind these effects.
Basic explanations for the electron anisotropy effects on the PFHI are offered 
by \cite{Kennel1968} and later by \cite{Michno2014}, namely, that for isotropic 
electrons the protons are weakly resonant, while for anisotropic electrons with 
$A_e>~1$ the phase velocity is increased and, consequently, the protons become less 
resonant leading to lower growth rates of PFHI. 
This explanation is confirmed here by studying the resonance conditions 
$\vert\xi_p^+\vert=~\vert\ (\tilde{\omega}+1)/(\tilde{k}\sqrt{\beta_{p,\parallel}})\vert$ 
given by the arguments of plasma dispersion 
function in Appendix~A, eq.~(7). We call these quantities resonant factors, and 
Figures \ref{f3}~(a) and (b) display them for protons and electrons, respectively, 
for the same plasma parameters as in Figure~\ref{f1}~(a). The zoomed plot in 
Figure~\ref{f3}~(a) provides details on the proton resonant factor $\vert\xi_p^+\vert$ for 
different values of $\delta$-index at wavenumbers $\tilde{k}$ corresponding to the 
peaks of the PFHI growth rates as $(\delta,\tilde{k}, \vert\xi_p^+\vert)=~(-0.37,0.42, 2.15)$, 
$(0.0,0.41,1.98)$, $(1.0,0.5,1.6)$. It becomes now clear that in the presence of 
anisotropic electrons with $A_e>~1$, the resonant factor increases $(\vert\xi_p^+\vert=2.15 > 1)$
and the protons become less resonant with the resulting PFHI. 
 \begin{figure}[t]
   \centering
      \includegraphics[width=7.2cm]{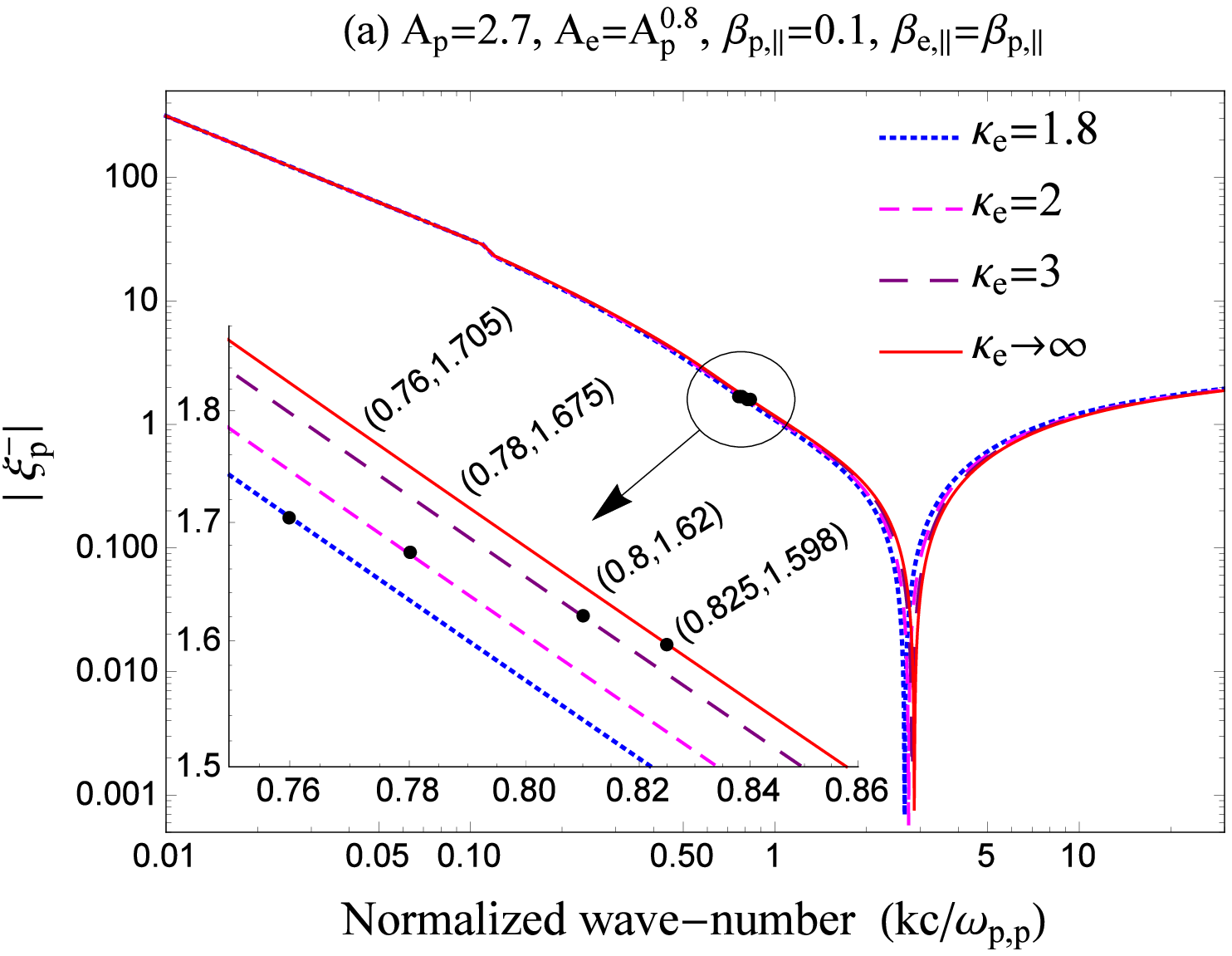}
   \includegraphics[width=7.2cm]{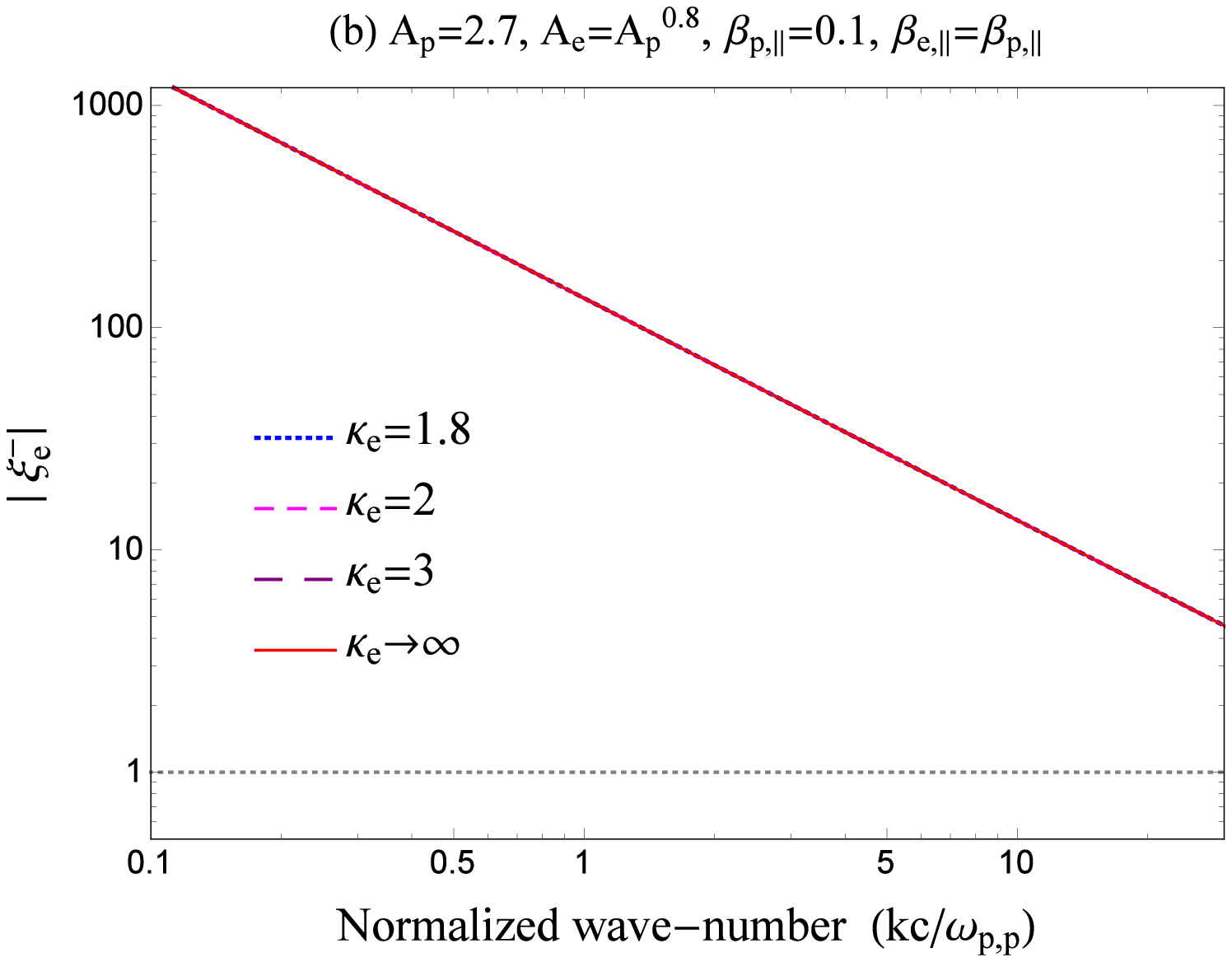}
      \caption{The resonant conditions for protons $\vert\xi_p^-\vert$ (top) and electrons $\vert\xi_e^-\vert$ (bottom) for the same plasma parameters in Figure (\ref{f2}-b).
              }
         \label{f4}
   \end{figure}
With increasing the wave-number the resonant factors $\vert\xi_p^+\vert$ drops down
to a minimum value, and then, for large enough $\tilde{k}>1$ their values rise again since the 
real frequency (not shown here) is rapidly increased. For $\delta=0.0,-0.37$ 
these minimum values remain above unity.
Comparison can be made to a simplified approach with isotropic electrons, i.e., $\delta=0.0$.
For a negative correlation-index $\delta=-0.37$ the protons become less resonant 
($\vert\xi_p^+\vert>1$) due to a significant increase of the real frequency by coupling to the 
high-frequency WI with the same RH polarization, see also Figure~\ref{f1}~(a). 
As normalized quantities, both the wave-frequency and the 
resonant factor remain much larger than unity with increasing the wave-number. In the opposite case, namely, for a positive 
correlation-index $\delta=1.0$, the EFHI arises at (slightly) higher wavenumbers, Figure~\ref{f2}~(a), with 
an opposite (LH) polarization, i.e., the wave-frequency changes the sign \citep{Michno2014}. 
The turning point (singularity given by the cold-plasma resonance $\Re(\xi_p^+) =0$), where
the resonant factor takes a minimum value $\vert\xi_p^+\vert_{min}=\vert\tilde{\gamma}/ (\tilde{k} 
\sqrt{\beta_{p,\parallel}})\vert\ll 1$, is followed by the resonance of protons with the EFHI,
i.e., $\vert\xi_p^+\vert \sim 1$, and with increasing the wave-number the resonant factor increases again. 

Figure~\ref{f3}~(b) shows the electron resonant factor $\vert\xi_e^+\vert$ at wave-numbers corresponding 
to the peaks of the EFHI and WI as $(\delta,\tilde{k}, \vert\xi_e^+\vert)=$(1.0,6.85,3.6),(-0.37,18.5, 1.06). 
The electrons are clearly non-resonant with the EFHI since $\vert\xi_e^+\vert>~1$, and strongly non-resonant 
near the peak of the PFHI where $\vert\xi_e^+\vert\gg~1$. However, the electrons become resonant with the 
WI when $\vert\xi_e^+\vert\gtrsim~1$. 

Additional insights can be provided for the EMIC instability in order to understand the 
recent results which show that anisotropic electrons with $A_e>1$ may increase the wave-frequency 
of these modes but inhibit the growth rates of the instability \citep{Shaaban2015}, while 
these effects are stimulated by the suprathermal electrons \citep{Shaaban2016Supra}. Our dynamical 
model which correlates the main kinetic properties of protons and electrons reconfirms
these effects, e.g., in Figure~\ref{f2}.
The resonant factors for protons ($\vert\xi_p^-\vert$) and electrons 
($\vert\xi_e^-\vert$) are plotted in Figure~\ref{f4}, panels (a) and (b), respectively. 
These factors are computed for the same parameters as in Figure~\ref{f2}~(b) to show 
the influence of suprathermal electrons quantified by the power-index $\kappa=1.8, 
2,3,\infty$. For protons, four values of the resonant factor are explicitly given
corresponding to the peaks of the EMIC growth rates as $(\kappa_e, \tilde{k},\vert\xi_p^-\vert)
=~(1.8,1.705)$, $(0.78,1.675)$, $(0.81,1.62)$, $(0.825,1.598)$. According to the zoomed plot 
in Figure~\ref{f4}~(a), the protons become less resonant $(\vert\xi_p^-\vert=1.705)$ with 
increasing the suprathermal population of electrons $(\kappa_e=1.8)$, which explains the 
inhibiting effect of the EMIC instability. Note that maximum growth rate for $\kappa_e=3$ 
is near the threshold level $\gamma_m/\Omega_p=10^{-2}$, and the corresponding value of 
the proton resonant factor $\vert\xi_p^-\vert=1.62$ is in a good agreement with the results 
obtained by \cite{Gary1994Correlation}. The proton resonant factor $\vert\xi_p^-\vert=~\vert\ 
(\tilde{\omega}-1)/(\tilde{k}\sqrt{\beta_{p,\parallel}})\vert$ decreases with increasing the 
wave-number until is drops abruptly down reaching a minimum value $\vert\xi_p^-\vert_{min}=
\vert\tilde{\gamma}/(\tilde{k}\sqrt{\beta_{p,\parallel}})\vert\ll 1$ at the turning point 
(singularity of cold-plasma resonance $\Re(\xi_p^-) =0$). Beyond this point the real 
frequencies are saturated and the modes are strongly damped with an increasing damping rate $-\tilde{\gamma}$, 
making the proton resonant factor (as absolute value) to increase.

Figure~\ref{f4}~(b) shows that the electrons are highly non-resonant near the peaks of the 
EMIC instability,  with $\vert\xi_e^-\vert\gg1$, and remain non-resonant $\vert\xi_e^-\vert\simeq10$, 
even for the wave-numbers corresponding to the electron scales. It becomes also clear that the EMIC branch
cannot connect to the high-frequency whistler (electron cyclotron) modes, which have a different (RH) 
polarization and which is resonantly destabilized by the (anisotropic) electrons, e.g., the 
whistler instability (WI) discussed above, also known as the electron-cyclotron instability.

 \begin{figure}[t]
   \centering
      \includegraphics[width=8cm]{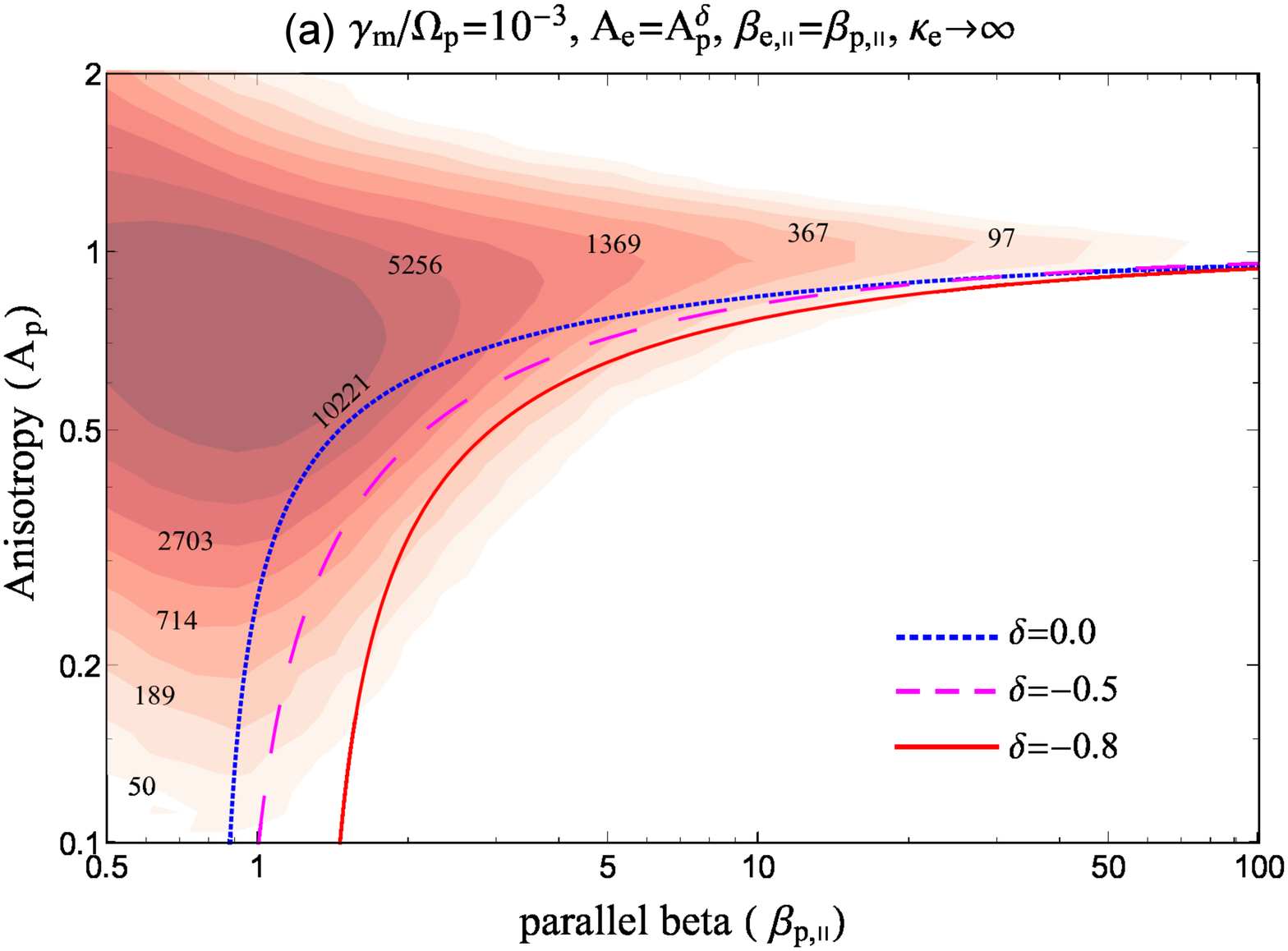}
      \includegraphics[width=8cm]{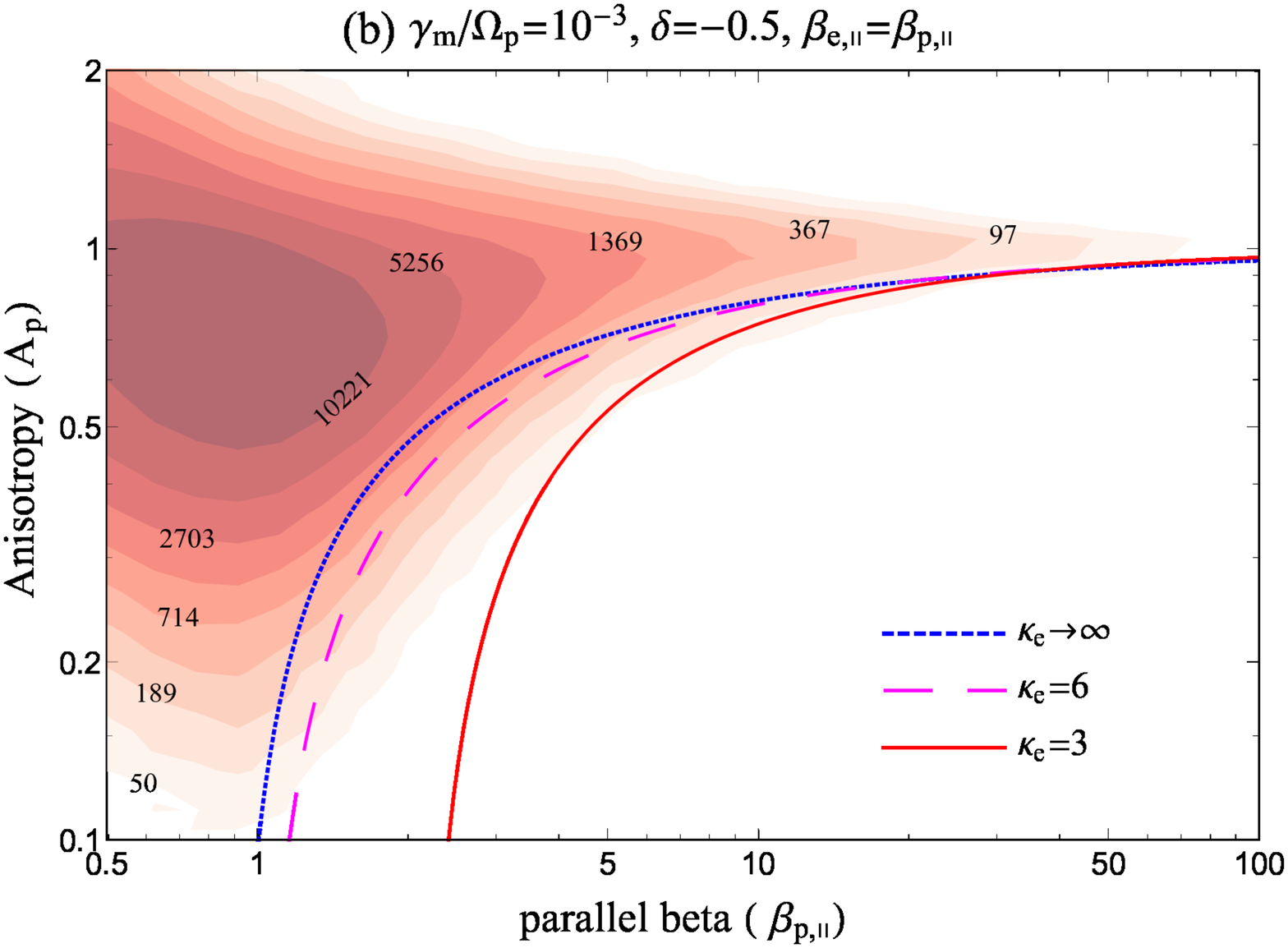}
      \caption{The influence of the correlation-index $\delta=0.0$, $-0.5,-0.8$ (top), and the power-index $\kappa=3,6,\infty$ with $\delta=-0.5$ (bottom) on the PFHI thresholds ($\gamma_m =10^{-3} \Omega_p$). Thresholds are compared with the proton (core) anisotropy at 1~AU in the solar wind and the plasma parameters are explicitly given in each panel
              }
         \label{f5}
   \end{figure}

\section{Instabilities thresholds vs. observations }

The anisotropy thresholds may provide a straightforward confirmation for the constraining 
role played by the kinetic instability in collision-poor plasmas from space, namely, when 
these thresholds fit the limits of the temperature anisotropy reported by the observations.
Derived for different levels of maximum growth-rates $\gamma_m/\Omega_p=~10^{-3}, 10^{-2}, 
10^{-1}$ and for an extended range of the plasma beta parameter $0.005\leqslant~
\beta_{p,\parallel}\leqslant~100$, the instability thresholds may also provide a general 
picture of the instability and the new effects triggered by the interplay of electrons 
and protons. In the present work we analyze the isocontours of anisotropy thresholds ($A_{p}$)
derived for $\gamma_m=~10^{-3}\Omega_p$ (a sufficiently low level also adopted in similar 
investigations), and represented as an inverse correlation law of the proton plasma beta 
$\beta_{p,\parallel}$ \citep{Hellinger2006}
\begin{equation}
A_{p}=1+\frac{a}{\left(\beta _{p,\parallel}-\beta_0\right)^{b}}. \label{e6}
\end{equation}
For the instability thresholds derived in Figures~\ref{f5}-\ref{f7} fitting parameters 
$a$, $b$, and $\beta_0$ are tabulated in tables \ref{t-1} and \ref{t-2} in 
Appendix B. The standard inverse correlation introduced by \citep{Gary1994Correlation} 
may be recovered for $\beta_0=0$. 
Thresholds are compared with the observations in the slow solar wind 
($v_{sw}\leqslant~600$~$km/s$), i.e., protons measured by SWE \citep{Ogilvie1995} 
and MFI \citep{Lepping1995} on the WIND spacecraft at 1~AU \citep{Kasper2002, Hellinger2006, 
Michno2014}.

 \begin{figure}[t]
   \centering
      \includegraphics[width=8.4cm]{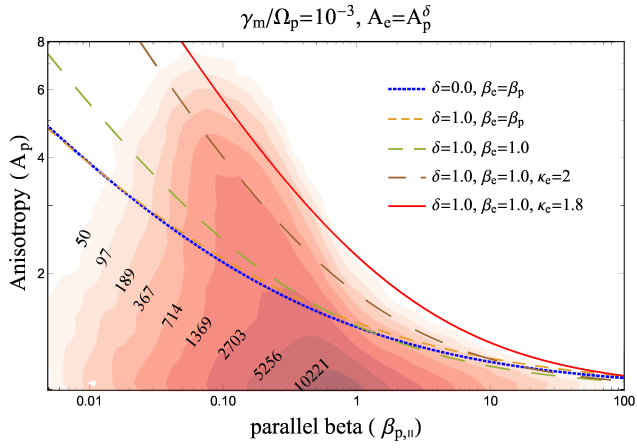}
      \caption{Thresholds conditions ($\gamma_m =10^{-3} \Omega_p$) for the EMIC instability compared with the proton (core) anisotropy at 1~AU in the solar wind. The plasma parameters are explicitly given in each panel.
              }
         \label{f6}
   \end{figure}

The PFHI thresholds are displayed in Figure~\ref{f5}, in panel~(a) for different correlation-indices 
$\delta=~0.0, -0.5$, $-0.8$ (implying different electron anisotropy $A_e=~A_p^{\delta}$), and in panel~(b)
for different power-indices $\kappa_e=3, 6, \infty$ and same $\delta=-0.5$. Simplified approaches usually 
adopt $\delta=0.0$ when the electrons are assumed isotropic $A_e=1$ (blue dotted line in Figure 5~(a)). The inhibiting 
effect obtained for a negative correlation index $\delta<0$ (i.e., $A_e>1$), see Figure~\ref{f1}~(a), 
is confirmed by the thresholds with $\delta=~-0.5,-0.8$ moving towards higher values of the proton plasma 
beta $\beta_{p,\parallel}$. By increasing this anti-correlation between protons and electrons, i.e., 
decreasing negative values of $\delta$, the instability thresholds are enhanced and can markedly 
improve their fit with the limits of the temperature anisotropy observed in the solar wind. 
In the second panel~(b) we can observe that these thresholds are further boosted by the suprathermal 
electrons: for lower values of $\kappa_e$ the PFHI thresholds are moved to higher plasma beta exceeding 
the limits observed for the proton anisotropy confirming the results in Figure~\ref{f1}~(b).  

Figure~\ref{f6} shows the influence of the anisotropic electrons with $A_e>1$ on the EMIC instability, 
and compares to the idealized case with isotropic electrons ($A_e=1$, dotted blue line). 
For $\beta_{e,\parallel}=\beta_{p,\parallel}$ the instability 
thresholds do not change much to improve fitting to the observations, even for $\delta=1$ 
(i.e., $A_p=A_e$), see the short-dashed (orange) line. 
However, \cite{Shaaban2015, Shaaban2016Supra} have recently shown that inhibiting effect induced by  
the anisotropic electrons is stimulated by increasing $\beta_{e,\parallel}$. 
Thus, for an average value $\beta_{e,\parallel}=1$ indicated by the observations \citep{Stverak2008}, 
also commonly invoked in similar investigations \citep{Hellinger2006, Matteini2013}, the EMIC threshold 
can increase considerably constraining more observational data, see dashed (green) line. 
On the other hand, the fit with the observations is considerably improved in the presence of 
suprathermal electrons, i.e., the instability thresholds enhance with decreasing the power-index 
$\kappa_e$ confirming the results in Figure~\ref{f2}. In this case, the EMIC thresholds are plotted for 
$A_e=A_p$ (i.e., $\delta=1$), $\beta_{e,\parallel}=1$, $\kappa_e=2$ (long-dashed brown line), and 
$\kappa_e=1.8$ (solid red line). 

 \begin{figure}[t]
   \centering
   \includegraphics[width=8cm]{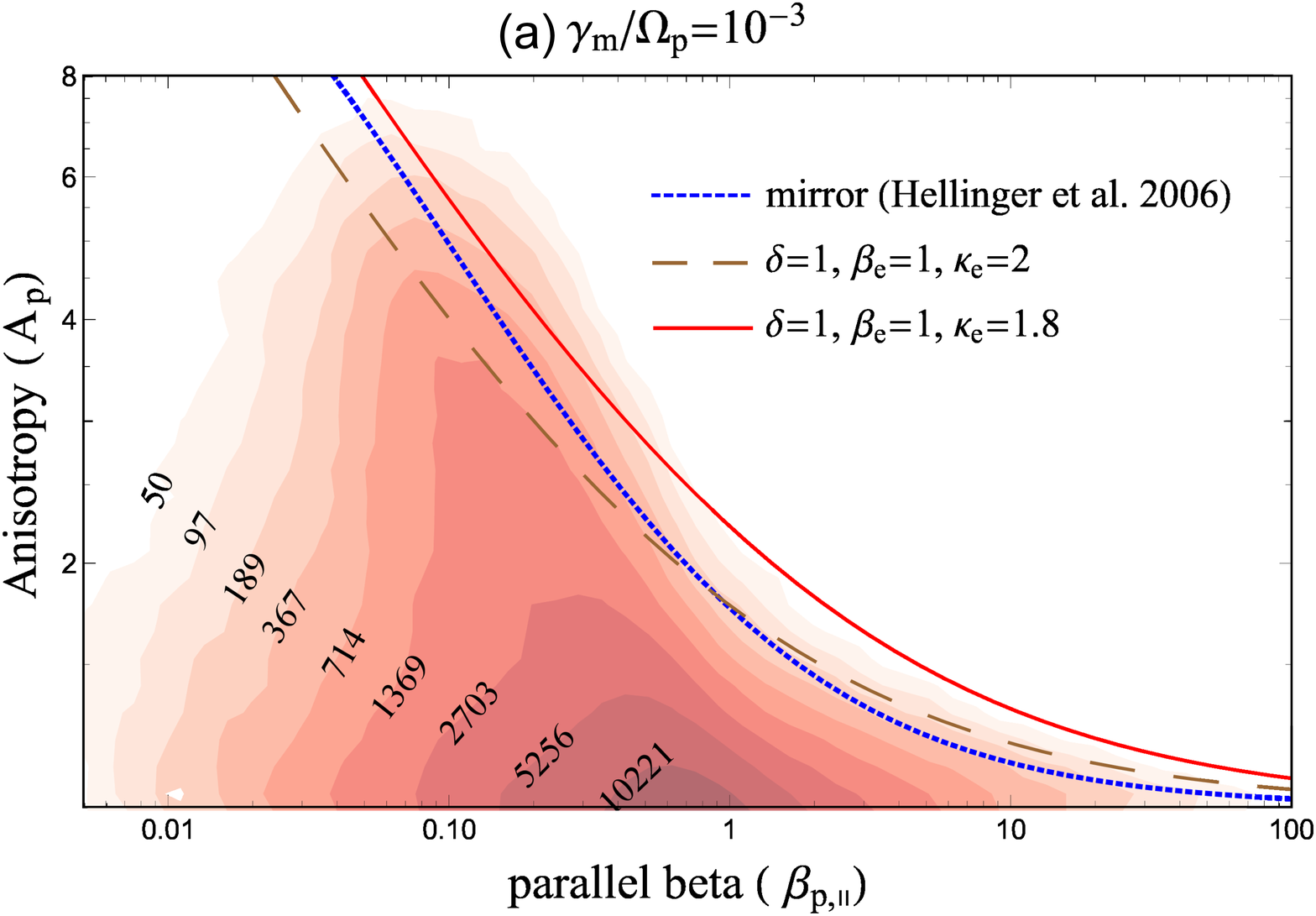}
    \includegraphics[width=8cm]{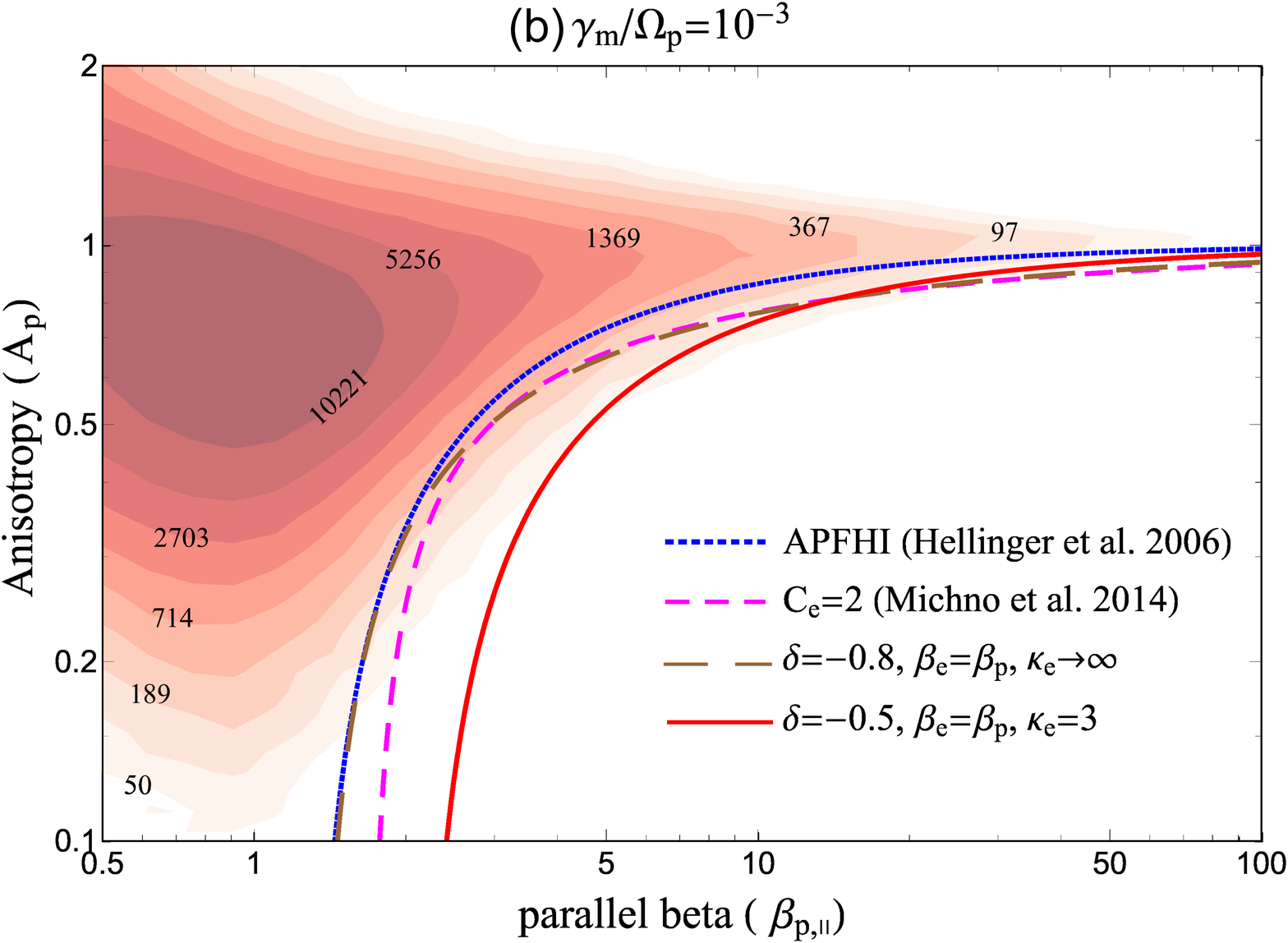}
      \caption{Thresholds of the aperiodic ($\omega_r = 0$) instabilities, i.e., mirror and the aperiodic PFH  
			(APFH), as provided by Hellinger et al. 2006 , and the instabilities of our periodic modes, i.e., 
			EMIC and PFH, are compared with the proton core data at 1 AU in the solar wind.
              }
         \label{f7}
   \end{figure}

For the sake of completeness, Figure (\ref{f7}) compares the best of our results 
with the those obtained by \cite{Michno2014} for parallel PFHI ($C_e=2$), and 
\cite{Hellinger2006} for the aperiodic $(\omega_r=0)$ instabilities, namely, the 
aperiodic proton firehose (APFH) and mirror instabilities. 
Most interesting appear to be the new instability thresholds obtained in panel~(a) for the EMIC instability
in the presence of suprathermal electrons and with a direct (positive) correlation $\delta = 1$ 
between the proton and electron anisotropies ($A_e=A_p$). These isocontours show a good alignment to the 
observations, similar to the mirror instability. \textbf{The presence of suprathermal electrons, which 
are ubiquitous in the solar wind, is critical and may considerably enhance the role played by the
EMIC instability in the low-beta regimes (e.g., for $\kappa_e=1.8$). Moreover, these results
are obtained for a common (average) value of the electron beta parameter, namely, $\beta_{e,\parallel} =1$ 
indicated by the observational data cumulation \citep{Stverak2008}, and for a direct correlation of 
the electron and proton anisotropies given by a positive $\delta = 1 >0$. As already explained 
in the Introduction we do not dispose of systematic observational analyses to confirm a direct 
correlation of electron and proton anisotropies, but this condition may in general be ensured by 
the mechanisms always at work in space plasmas, e.g., solar wind expansion, magnetic focusing, 
and it seems therefore more plausible than an anti-correlation of the anisotropies. Based on 
these arguments, the fits obtained in panel~(a) can be considered robust enough to support the implication of 
the EMIC instability in constraining the proton temperature anisotropy and explain the observations.}
In panel~(b) we compare the new thresholds obtained for the PFHI  with those derived by \cite{Michno2014} 
or \cite{Hellinger2006} for the APFH instability, and can conclude that anisotropic electrons may influence
the PFHI thresholds and determine them to align better to the observations, e.g., the long-dashed brown line 
obtained for $\delta=-0.8$. However, in this case the best-aligned thresholds are conditioned by an anti-correlation 
between the proton and electron anisotropies as required by a negative $\delta < 0$.

\section{Conclusions}
In a collisionless electron-proton plasma the temperature anisotropy, and, implicitly, 
the distribution function should be regulated by the resulting instabilities and 
electromagnetic fluctuations through the wave-particle interaction. For anisotropic 
protons, the  theory and simulations predict a dominance of the cyclotron modes driven 
unstable by the PFH and EMIC instabilities, which may develop fast enough leading to 
an important pitch angle scattering of protons toward isotropy 
\citep{Kennel1968, Gary1992}. However, predictions made by the simplified approaches 
for the anisotropy thresholds of these instabilities appear to be overestimated 
by comparison to the temperature anisotropy measured in-situ in the solar wind. Instead, 
these observations seem to be better constrained by the aperiodic instabilities, e.g., 
APFHI and mirror instability \citep{Hellinger2006}. 

In this paper we have aimed to resolve this paradigm and provide more realistic 
predictions from an advanced modeling that accounts for the interplay of protons and electrons, 
and the presence of suprathermal electrons. 
New regimes are thus found for the PFH and EMIC instabilities, which are mainly 
controlled by the cumulative effects of protons and electrons by correlating
either their anisotropies, e.g., $A_e=A_p^\delta$, via the parameter $\delta$, or/and their 
plasma beta parameters, $\beta_e=\beta_p$. Similarities and differences among these regimes 
are highlighted in Section~3, always comparing with the idealized approaches which consider
electrons isotropic, i.e., $\delta = 0$. We have studied the effects of these correlation
factors on the growth-rates and provided physical explanations by studying the resonant 
conditions for both the protons and electrons. 

A comparative analysis of these new regimes enabled us to identify conditions that may 
inhibit the instabilities and make their thresholds to adjust better to the observations. 
Thus, the PFHI is driven by a proton anisotropy $A_p<1$, and can be inhibited by anisotropic
electrons with anti-correlated anisotropies $A_e=A_p^\delta > 1$ given by a negative 
$\delta < 0$. In the opposite situation when protons exhibit a temperature anisotropy $A_p>1$, 
the EMIC instability is inhibited only for a positive $\delta > 0$ meaning electrons with 
a direct correlated anisotropy $A_e=A_p^\delta > 1$. Moreover, in both these two cases 
the inhibiting effect is boosted by the suprathermal electrons. The explanation is provided
by the resonant factors, which indicate that protons become less resonant inhibiting the 
instability and leading to higher anisotropy thresholds. These results confirm the expectations 
from the previous studies carried out by \cite{Kennel1968, Lazar2011, Michno2014, Shaaban2015, Shaaban2016Supra}.

To provide a complete picture, in Section~4 we have studied the instability thresholds, and recovered
the inhibiting effects on the PFH and EMIC instabilities for an extended range of the plasma beta parameter 
$0.005 <\beta_{p,\parallel} < 100$. These thresholds are compared to the proton anisotropy data 
observed in the (slow) solar wind at 1~AU. The PFHI thresholds decrease moving towards higher plasma 
beta $\beta_{p,\parallel}$ with decreasing the $\delta-$index and with increasing the 
suprathermal electron population.
For the EMIC instability, the threshold conditions in the low-beta regimes are only weakly affected by 
the anisotropic electrons, but can be significantly lowered by increasing the electron plasma beta 
$\beta_{e,\parallel}=1$ and the presence of suprathermal electrons. 
To conclude, we have identified the conditions for the instability thresholds to align and shape the limits
of the temperature anisotropy reported by the observations. These agreements with the observations
can be even better than those obtained before for the aperiodic instabilities, but are highly 
conditioned by the electron properties, i.e., the anisotropy (correlated or anti-correlated
with the proton anisotropy), plasma beta parameter, and their suprathermal populations.

Suprathermal electrons are ubiquitous in space plasmas but the main question arising now concerns the 
existence of the plasma states with protons and electrons having anisotropies either direct correlated 
and roughly described by a positive $\delta > 0$, or anti-correlated by a negative $\delta < 0$. 
As already discussed in the Introduction, we do not dispose of any systematic
evidences and estimations of these correlations  from the observations, but there are some qualitative 
indications which appear to be more favorable to direct correlated 
anisotropies of electrons and protons, i.e., $A_{e,p} > 1$ at low heliocentric distances
for both species, or $A_{e,p} < 1$ at higher radial distances, again for both species. These 
states may naturally result from the large-scale processes which generate temperature anisotropy
in the solar wind, e.g., adiabatic expansion, magnetic field compression, while the other states 
with anti-correlated anisotropies may be established most probably, locally, by the small-scale 
mechanisms of particle energization involving the electromagnetic fluctuations. 
For a correct interpretation of the kinetic effects, like dissipation or instabilities, in space 
plasmas, a self-consistent treatment of plasma particles and electromagnetic 
fluctuations is therefore crucial and needs to be supported by the observations. From this point
of view, the advanced approach proposed in the present paper may be considered as an important 
progress towards a realistic interpretation of the interplay of eletrons and protons and their
effects in the solar wind.

\begin{acknowledgements}
\textbf{Acknowledgements} The authors acknowledge the use of WIND
SWE (Ogilvie et al. 1995) ion data, and WIND MFI (Lepping
et al.\ 1995) magnetic field data from the SPDF CDAWeb service:
http://cdaweb.gsfc.    nasa.gov/. The authors acknowledge support
from the Katholieke Universiteit Leuven. These results were obtained
in the framework of the projects GOA/2015-014 (KU Leuven),
G.0A23.16N (FWO-Vlaanderen), and C~90347 (ESA Prodex). The
research leading to these results has also received funding from IAP
P7/08 CHARM (Belspo). S.M. Shaaban would like to thank the
Egyptian Ministry of Higher Education for supporting his research activities.
\end{acknowledgements}

\section*{Appendix A: Distributions and dispersion functions}
For a plasma of electrons and protons\;  with\;   bi-Maxwellian velocity distribution functions (VDFs)
\begin{equation}
F_{\alpha,M}\left( v_{\parallel },v_{\perp }\right) =\frac{1}{\pi
^{3/2}u_{\alpha ,\perp }^{2} u_{\alpha ,\parallel }}\exp \left(
-\frac{v_{\parallel }^{2}} {u_{\alpha,\parallel }^{2}}-\frac{v_{\perp
}^{2}}{u_{\alpha ,\perp }^{2}}\right),  \label{e4}
\end{equation}
where thermal velocities $u_{\alpha,\parallel, \perp}$  are defined by the components of the anisotropic temperature
\begin{equation}
T_{\alpha,\parallel}^M=\frac{m}{k_B}\int d\textbf{v} v_\parallel^2
F_\alpha(v_\parallel, v_\perp)=\frac{m u_{\alpha ,\parallel}^2}{2 k_B}
\end{equation}
\begin{equation}
T_{\alpha,\perp}^M=\frac{m}{2 k_B}\int d\textbf{v} v_\perp^2
F_\alpha(v_\parallel, v_\perp)=\frac{m u_{\alpha,\perp}^2}{2 k_B},
\end{equation}
the plasma dispersion function in Eq.(\ref{e1}) takes the standard form \citep{Fried1961}
\begin{equation}
Z_{\alpha,M}\left( \xi _{\alpha,M}^{\pm }\right) =\frac{1}{\pi ^{1/2}}\int_{-\infty
}^{\infty }\frac{\exp \left( -x^{2}\right) }{x-\xi _{\alpha,M}^{\pm }}dt,\
\ \Im \left( \xi _{\alpha,M}^{\pm }\right) >0  \label{e5}
\end{equation}
of argument $\; \xi _{\alpha,M}^{\pm }= \left(\omega \pm \Omega _{\alpha}\right)/\left(ku_{\alpha,\parallel,}\right)$

To include suprathermal population, the electrons can be described by a bi-Kappa 
VDF \citep{Summers1991}
\begin{equation}
     \begin{aligned}
F_{e,\kappa}= \frac{1}{\pi ^{3/2}u_{e, \perp }^{2}u_{e, \parallel}}&\frac{\Gamma\left(\kappa_e +1\right) }{\Gamma \left(\kappa_e -1/2\right) }\\
&\left[ 1+\frac{v_{\parallel }^{2}}{\kappa_e u_{e, \parallel }^{2}}+\frac{v_{\perp }^{2}}{\kappa_e u_{e, \perp }^{2}}\right] ^{-\kappa_e -1} \label{e7}
     \end{aligned}
\end{equation}
which is normalized to unity $\int d^{3}vF_{e,\kappa}=~1$, and is written in terms of 
thermal velocities $u_{e,\parallel, \perp}$ defined by the components of the effective 
temperature (for a power-index $\kappa_e >3/2$)
\begin{equation}
T_{e,\parallel}^K=\frac{2 \kappa_e}{2 \kappa_e-3}\frac{m_e u_{e ,\parallel}^2 }{2 k_B}, \;\ T_{e,\perp}^K=\frac{2 \kappa_e}{2 \kappa_e-3}\frac{m_e u_{e ,\perp}^2}{2
k_B}.  \label{e8}
\end{equation}
Suprathermals enhance the electron temperature, and implicitly the plasma beta parameter \citep{Leubner2000,Leubner2001,Lazar2015Destabilizing}
\begin{equation}
  \begin{aligned}
T_{e,\parallel, \perp}^K=\frac{2 \kappa_e}{2 \kappa_e-3}T_{e,\parallel,\perp}^M
> T_{e,\parallel,\perp}^M, \\\beta_{e,\parallel, \perp}^K=\frac{2 \kappa_e}{2 \kappa_e-3}
\beta_{e,\parallel,\perp} > \beta_{e,\parallel,\perp},
  \end{aligned}  \label{e9}
\end{equation}
and for the modified Kappa dispersion function (\ref{e5}) we use in Eq.(\ref{e1}) the form \citep{Lazar2008}
\begin{equation}
     \begin{aligned}
     Z_{e,\kappa}\left(\xi_{e,\kappa}^{\pm }\right)&=\frac{1}{\pi ^{1/2}\kappa_{e}^{1/2}}\frac{\Gamma \left( \kappa_{e} \right) }{\Gamma \left(\kappa_{e} -1/2\right) }\\     
           &\times\int_{-\infty }^{\infty }\frac{\left(1+x^{2}/\kappa_{e} \right) ^{-\kappa_{e}}}{x-\xi_{e,\kappa}^{\pm }}dx,\ \Im \left(\xi _{e,\kappa}^{\pm }\right) >0,  \label{e10}
     \end{aligned}
\end{equation}
of argument $\;\xi_{e,\kappa}^{\pm }=\left(\omega \pm \Omega _{e}\right)/\left(k u_{e,\parallel,}\right).$

\section*{Appendix B: Fitting parameters for Eq.(\ref{e6})}

\begin{table}[h]
\caption{Fitting parameters for PFH thresholds in Figure~\ref{f5} and \ref{f7} (b).}            
\label{t-1}     
\centering                         
\begin{tabular}{c c c c c}       
\hline\hline                 
 $\kappa$ & $\delta$ & $a$ & $b$ & $\beta_0$ \\   
\hline                        
 $\infty$ &   0.0   & $-0.453$ & 0.467 & 0.652 \\      
 $\infty$ & $-0.5$  & $-0.733$ & 0.607 & 0.293 \\
 $\infty$ &  $-0.8$ & $-0.772$ & 0.543 & 0.709 \\
        2 &  $-0.5$ & $-1.849$ & 0.865 & 0.112 \\
        6 &  $-0.5$ & $-0.990$ & 0.707 & 0.0 \\
  
\hline                                   
\end{tabular}
\end{table}

\begin{table}[h]
\caption{Fitting parameters for EMIC thresholds in Figure~\ref{f6} and \ref{f7} (a).}            
\label{t-2}     
\centering                         
\begin{tabular}{c c c c c}       
\hline\hline                 
 $\kappa$ & $\delta$ & $\beta_{e,\parallel}$ & $a$ & $b$ \\   
\hline                        
   $\infty$ & 0 & $\beta_{p,\parallel}$ & 0.451 & 0.402 \\      
   $\infty$ & 1 & $\beta_{p,\parallel}$ & 0.493 & 0.382 \\
   $\infty$ & 1 &            1          & 0.463 & 0.495 \\
          2 & 1 &            1          & 0.774 & 0.590 \\
        1.8 & 1 &            1          & 1.221 & 0.579 \\ 
\hline                                   
\end{tabular}
\end{table}

\bibliographystyle{spr-mp-nameyear-cnd}
\bibliography{allpapers}
\end{document}